# A simple mathematical proof of Boltzmann's equal *a priori* probability hypothesis

Denis J. Evans,[1] Debra J. Searles[2] and Stephen R. Williams[1]

[1] Research School of Chemistry, Australian National University, Canberra, ACT 0200, Australia

[2] Nanoscale Science and Technology Centre, School of Biomolecular and Physical Sciences,

Griffith University, Brisbane, Qld 4111 Australia

**Abstract**

Using the Dissipation Theorem and a corollary of the Fluctuation Theorem, namely the Second Law Inequality, we give a first-principles derivation of Boltzmann's postulate of equal *a priori* probability in phase space for the microcanonical ensemble. We show that if the initial distribution *differs* from the uniform distribution over the energy hypersurface, then under very wide and commonly satisfied conditions, the initial distribution will relax to that uniform distribution. This result is somewhat analogous to the Boltzmann H-theorem but unlike that theorem, applies to dense fluids as well as dilute gases and also permits a nonmonotonic relaxation to equilibrium. We also prove that in ergodic systems the uniform (microcanonical) distribution is the only stationary, dissipationless distribution for the constant energy ensemble.

Most textbook discussions of the equilibrium microcanonical phase space distribution functions rely on Boltzmann's postulate of equal *a priori* probability in phase space [2-5]. Arguments are then given for various microscopic expressions for the various macroscopic thermodynamic quantities. No attempt is made to prove the Boltzmann's postulate.

A second approach by-passes the microcanonical ensemble [1-3] and seeks to propose a microscopic definition for the entropy in the canonical ensemble and then attempts to show that the standard canonical distribution function can be obtained by maximising the entropy subject to the constraints that the distribution function should be normalized and that the average energy is constant. The choice of the second constraint is completely subjective due to the fact that at equilibrium, the average of every phase function is fixed. Why single out the energy?

The relaxation of systems to equilibrium is also fraught with difficulties. The only reasonably general approach to this problem is summarized in the Boltzmann H-theorem. Beginning with the definition of the H-function, Boltzmann proved that the Boltzmann equation for the time evolution of the single particle probability density in an ideal gas, implies that in spatially uniform gases the H-function cannot increase [2, 4]. There are at least four problems with this. Firstly the Boltzmann equation is only valid for an ideal gas. Secondly and more problematically, unlike Newton's equations the Boltzmann equation itself is not time reversal symmetric so there is no surprise that one can prove relaxation. Thirdly the Boltzmann H-theorem only allows a monotonic relaxation to equilibrium. Everyday experience of the behaviour of dense fluids and solids shows that the process of relaxation to equilibrium is *not* necessarily monotonic. Lastly the H-theorem only applies to dilute gases that are spatially uniform.

Ergodic theory [6] has certainly given a derivation of the fact that if a system is *mixing* and *ergodic* then the system will eventually relax to a unique equilibrium state. However, ergodic theory is extremely technical and these derivations have not become available to textbook descriptions of statistical mechanics. To this day very few systems have been shown mathematically to be ergodic. It is believed that ergodicity is very common in naturally occuring systems. In the present letter unless stated otherwise, we assume our systems are ergodic.

Probably the most complete summary of this unsatisfactory state of affairs can be found in the famous encyclopedia article [7] of 1912. In this article the Ehrenfests advocate the axiomatic approach.

Recently a number of new exact fluctuation relations have been proven. The transient Fluctuation Theorem (FT) of Evans and Searles [8] has been derived and its predictions confirmed in laboratory experiments [9]. The FT is valid for arbitrary densities. Most importantly it can be derived using time reversible microscopic dynamics. The FT is closely related to the Crooks Fluctuation Theorem [10] and to the Jarzynski Equality [11, 12]. However to study the relaxation of a system towards equilibrium only the Evans-Searles FT is relevant.

The FT is remarkable in that it represents one of the few exact results that apply to nonequilibrium systems far from equilibrium. It provides a generalized form of the 2$^{nd}$ Law of Thermodynamics that applies to small systems observed for short periods of time [8]. The FT gave the first rigorous explanation of how irreversible macroscopic behaviour arises from time reversible deterministic dynamics and therefore resolves the long-standing Loschmidt paradox [7]. More recently it has been shown that the dissipation function which is the argument of the FT is also the argument of transient time correlation functions whose time integrals give exact expressions for the linear and nonlinear response [13] of many particle systems.

In the present Letter we show how the dissipation function, the FT and its corollary the Second Law Inequality [14] can be used to prove that under certain assumptions, the postulate of equal *a priori* probability in phase space for the microcanonical ensemble. We also show that if the distribution is not uniform, that such an ensemble of systems will on average relax to the uniform (microcanonical) distribution and that this relaxation process need not proceed monotonically.

Consider a classical system of N interacting particles in a volume V. The microscopic state of the system is represented by a phase space vector of the coordinates and momenta of all the particles, $\{\mathbf{q}_1, \mathbf{q}_2, ..\mathbf{q}_N, \mathbf{p}_1, ..\mathbf{p}_N\} \equiv (\mathbf{q}, \mathbf{p}) \equiv \mathbf{\Gamma}$ where $\mathbf{q}_i, \mathbf{p}_i$ are the position and momentum of particle i.

Initially (at t = 0), the microstates of the system are distributed according to a normalized probability distribution function $f(\Gamma,0)$. We assume this distribution is an even function of the momenta, $f(\mathbf{q},\mathbf{p},0) = f(\mathbf{q},-\mathbf{p},0)$. We write the equations of motion for the N-particle system as

$$\dot{\mathbf{q}}_i = \mathbf{p}_i / m_i$$
$$\dot{\mathbf{p}}_i = \mathbf{F}_i(\mathbf{q}) \qquad (1)$$

where $\mathbf{F}_i(\mathbf{q}) = -\partial \Phi(\mathbf{q})/\partial \mathbf{q}_i$ is the interatomic force on particle i, and $\Phi(\mathbf{q})$ is the interparticle potential energy. These equations of motion preserve the phase space volume, $\Lambda = (\partial/\partial \Gamma) \cdot \dot{\Gamma} = 0$ : a condition known as the adiabatic incompressibility of phase space, or AI$\Gamma$ [15]. They also conserve energy, $H_0(\Gamma) \equiv \sum p_i^2 / 2m + \Phi(\mathbf{q})$ and total momentum $\sum_{i=1}^{N} \mathbf{p}_i$.

The exact equation of motion for the N-particle distribution function is the time reversible Liouville equation,

$$\frac{df(\Gamma,t)}{dt} = 0 . \qquad (2)$$

The derivation of the Evans-Searles FT [8] considers the response of a system that is initially described by some phase space distribution. The initial distribution is not necessarily at equilibrium, however we assume $\sum \mathbf{p}_i = \mathbf{0}$ because all thermodynamic quantities must be Galilae invariant. We can write any such initial distribution in the form

$$f(\Gamma,0) = \frac{\delta(H_0 - E)\delta(\mathbf{p})\exp[-F(\Gamma)]}{\int d\Gamma \, \delta(H_0 - E)\delta(\mathbf{p})\exp[-F(\Gamma)]} , \qquad (3)$$

where $F(\Gamma)$ is a single valued real function which is even in the momenta and E is the fixed value of the internal energy of the system. Since the initial distribution is even in the momenta, the FT states that provided the system satisfies the condition of ergodic consistency [7], the dissipation function $\Omega(\Gamma)$, defined as

$$\int_0^t ds\, \Omega(\Gamma(s)) \equiv \ln\left(\frac{f(\Gamma(0),0)}{f(\Gamma(t),0)}\right) - \int_0^t \Lambda(\Gamma(s))\,ds$$

$$= \ln\left(\frac{f(\Gamma(0),0)}{f(\Gamma(t),0)}\right) \qquad , \qquad (4)$$

$$\equiv \bar{\Omega}_t(\Gamma(0))t$$

satisfies the following time reversal symmetry [8]:

$$\frac{p(\bar{\Omega}_t = A)}{p(\bar{\Omega}_t = -A)} = \exp[At] . \qquad (5)$$

The derivation of the FT is straightforward and has been reviewed a number of times [8].

It is important to remember that the existence of the dissipation function $\Omega(\Gamma)$ at a phase point $\Gamma$, requires that $f(\Gamma,0) \neq 0$. The existence of the integrated form of the dissipation function requires that the dynamics is ergodically consistent (i.e. $\forall \Gamma, t$ such that $f(\Gamma,0) \neq 0, f(\Gamma(t),0) \neq 0$). Since the energy is a constant of the motion, ergodic consistency requires that we only consider states on the $H_0(\Gamma) = E$ hypersurface. We do not consider the shell microcanonical ensemble of states within a narrow band of energies, nor do we consider the cumulative microcanonical ensemble of all states less than a set energy.

The existence of the dissipation function only requires that the initial distribution is normalizable, even in the momenta and that ergodic consistency holds. To prove (5) requires an additional condition: the dynamics must be time reversal symmetric.

The FT leads to a number of corollaries such as the Second Law Inequality [14],

$$\left\langle \bar{\Omega}_t \right\rangle_{f(\mathbf{\Gamma},0)} \geq 0, \quad \forall t, f(\mathbf{\Gamma},0), \tag{6}$$

and the NonEquilibrium Partition Identity [17]), i.e. $\left\langle e^{-\bar{\Omega}_t t} \right\rangle = 1, \quad \forall t$. The notation $\left\langle ... \right\rangle_{f(\mathbf{\Gamma},0)}$ implies that the ensemble average is taken over the ensemble defined by the initial distribution $f(\mathbf{\Gamma},0)$, Eq. (3).

We have recently derived the Dissipation Theorem [13], a generalization of response theory to handle arbitrary initial distributions, which shows that, as well as being the subject of the FT, the dissipation function is also the central argument of both linear (i.e. Green-Kubo theory) and nonlinear response theory. The Dissipation Theorem allows the treatment of systems that are driven away from equilibrium by an external field or, as in this paper, systems that have no externally applied field but that are initially in nonequilibrium distributions. For the dynamics considered here, (1), the solution of the Liouville equation (2) can be written as,

$$f(\mathbf{\Gamma}(t),t) = f(\mathbf{\Gamma}(0),0). \tag{7}$$

From the definition of the dissipation function, (4),

$$f(\mathbf{\Gamma}(0),0) = f(\mathbf{\Gamma}(t),0) e^{\int_0^t ds\, \Omega(s)}. \tag{8}$$

Substituting (7) into (8) gives

$$f(\mathbf{\Gamma}(t),t) = e^{\int_0^t ds\, \Omega(\mathbf{\Gamma}(s))} f(\mathbf{\Gamma}(t),0), \forall \mathbf{\Gamma}(t). \tag{9}$$

Realising that $\mathbf{\Gamma}(t)$ is just a dummy variable,

$$f(\mathbf{\Gamma}(0),t) = e^{\int_0^t ds\, \Omega(\mathbf{\Gamma}(s-t))} f(\mathbf{\Gamma}(0),0)$$

$$= e^{-\int_0^{-t} d\tau\, \Omega(\mathbf{\Gamma}(\tau))} f(\mathbf{\Gamma}(0),0) \quad . \tag{10}$$

Thus the propagator for the N-particle distribution function has a very simple relation to exponential time integral of the dissipation function. By forming time averages of an arbitrary phase variable $B(\mathbf{\Gamma})$ using (10), and then differentiating and re-integrating with respect to time [13], we can write time averages of an arbitrary phase function $B(\mathbf{\Gamma})$ as

$$\langle B(\mathbf{\Gamma}(t)) \rangle_{f(\mathbf{\Gamma}(0),0)} = \langle B(\mathbf{\Gamma}(0)) \rangle_{f(\mathbf{\Gamma}(0),0)} + \int_0^t ds\, \langle \Omega(\mathbf{\Gamma}(0)) B(\mathbf{\Gamma}(s)) ] \rangle_{f(\mathbf{\Gamma}(0),0)} . \tag{11}$$

The derivation of Eq. (11) is called a Dissipation Theorem. We have derived it for the case where the system is initially in an arbitrary state and the dynamics is simply Newtonian. Versions of this Theorem are extremely general and can be derived for systems that are driven away from equilibrium by external fields. Like the FT and (11) all the various versions of the Dissipation Theorem are valid arbitrarily far from equilibrium.

For the Newtonian equations of motion (1) that conserve energy and momentum, consider the initial distribution:

$$f(\mathbf{\Gamma},0) \equiv f_{\mu C}(\mathbf{\Gamma},0) = \frac{\delta(\mathbf{p})\delta(H_0(\mathbf{\Gamma})-E)}{\iint d\mathbf{\Gamma}\; \delta(\mathbf{p})\delta(H_0(\mathbf{\Gamma})-E)} , \tag{12}$$

where $H_0(\mathbf{\Gamma})$ is the internal energy of the system.

It is trivial to show that for this distribution and the dynamics (1), the dissipation function $\Omega_{\mu C}(\mathbf{\Gamma})$, is identically zero

$$\Omega_{\mu C}(\mathbf{\Gamma}) = 0, \quad \forall \mathbf{\Gamma} \tag{13}$$

everywhere on the energy hypersurface, and from (10). we see that this initial distribution is preserved everywhere in the accessible phase space.

$$f(\Gamma,t) = f_{\mu C}(\Gamma,0), \forall \Gamma, t. \tag{14}$$

For ergodic systems we call distributions that are time independent and dissipationless, *equilibrium distributions*. We shall call the distribution given in (12), the microcanonical distribution.

Now consider an *arbitrary* deviation from the *microcanonical distribution*

$$f(\Gamma,0) = \frac{\delta(\mathbf{p})\delta(H_0(\Gamma)-E)\exp[-\gamma g(\Gamma)]}{\int d\Gamma\,\delta(\mathbf{p})\delta(H_0(\Gamma)-E)\exp[-\gamma g(\Gamma)]}. \tag{15}$$

where $g(\Gamma)$ is also even in the momenta. The factor $\gamma$ is a positive scaling parameter that can be used (if required) to scale the magnitude of deviations from the microcanonical distribution. For such a system evolving under our dynamics, the time integrated dissipation function is

$$\overline{\Omega}_t(\Gamma(0))t = \gamma[g(\Gamma(t)) - g(\Gamma(0))] \equiv \gamma\Delta g(\Gamma(0),t) \tag{16}$$

and (10) becomes

$$f(\Gamma,t) = \exp[-\gamma\Delta g(\Gamma,-t)]f(\Gamma,0). \tag{17}$$

Thus, if g is not a constant of the motion, there is dissipation and furthermore the distribution function will not be preserved. Therefore if the system is ergodic the only possible unchanging distribution function will be that where $\gamma g(\Gamma)$ is constant and in this case the distribution reduces to the equilibrium microcanonical distribution function.

If the system is not ergodic the phase space breaks up into non-overlapping ergodic subspaces labelled $D_\alpha$ and $S_\alpha(\Gamma) = 1$ if $\Gamma \in D_\alpha; = 0$ otherwise. The set of phase functions $\{S_\alpha(\Gamma); \alpha = 1, N_D\}$, where $N_D$ is the total number of ergodic sub-domains, comprise a set of constants of the motion. Within each sub-domain any deviation from local microcanonical weighting will induce dissipation and the distribution will not be preserved. The only possible unchanging distribution is microcanonical within each domain. The inter-domain weights are arbitrary however. Thus we have a new derivation of the Williams-Evans quasi-equilibrium distribution function for glassy systems[19].

Further, the dissipation function satisfies the Second Law Inequality ,

$$\gamma \langle \Delta g(\mathbf{\Gamma},t) \rangle_{f(\mathbf{\Gamma},0)} = \int_0^\infty A(1-e^{-A})p(\gamma \Delta g(\mathbf{\Gamma},t) = A)dA$$
$$\geq 0, \quad \forall t, f(\mathbf{\Gamma},0)$$
(18)

which can only take on a value of zero when $\gamma \Delta g(\mathbf{\Gamma},t) = 0, \forall \mathbf{\Gamma}$. Thus if the system is ergodic and the initial distribution differs from the microcanonical distribution, (12), there will *always* be dissipation and on *average* this dissipation is *positive*. This remarkable result is true for arbitrary $\gamma$. Thus for ergodic systems we have derived an expression for the unique constant energy equilibrium state and shown that it takes on the standard form for the microcanonical distribution, modulo the facts that the momentum and energy are constants of the motion. If the system is not ergodic then the results apply within each ergodic subdomain. The weights between the various subdomains are history dependent and arbitrary. The weights are only uniform within the ergodic sub-domains. The full non-ergodic equilibrium distribution is therefore not unique.

This completes our first-principles derivation of the equilibrium distribution function. We now consider the question of relaxation towards equilibrium. We assume the system is ergodic. Substituting (16) into (11) and choosing the arbitrary phase function, B, in (11) to be g itself gives:

$$\langle g(t) \rangle_{f(\mathbf{\Gamma},0)} = \langle g(0) \rangle_{f(\mathbf{\Gamma},0)} + \gamma \int_0^t ds \langle \dot{g}(0)g(s) \rangle_{f(\mathbf{\Gamma},0)} \geq 0, \quad \forall t ,$$
(19)

where the ensemble averages are with respect to the distribution function (15).

We assume that at sufficiently a long time the correlations vanish so that for t > $t_c$, where $t_c$ is the correlation time, we can write :

$$\begin{aligned}\langle g(t)\rangle_{f(\Gamma,0)} &= \langle g(0)\rangle_{f(\Gamma,0)} + \gamma\int_0^{t_c} ds\langle \dot{g}(0)g(s)\rangle_{f(\Gamma,0)} + \gamma\int_{t_c}^{t} ds\langle \dot{g}(0)\rangle_{f(\Gamma,0)}\langle g(s)\rangle_{f(\Gamma,0)}\\ &= \langle g(0)\rangle_{f(\Gamma,0)} + \gamma\int_0^{t_c} ds\langle \dot{g}(0)g(s)\rangle_{f(\Gamma,0)}\end{aligned} \quad (20)$$

The final result is obtained after noting g is an even function of time. This means for t > $t_c$, the value $\langle g(t)\rangle_{f(\Gamma,0)}$ does not change with time, or $\langle \dot{g}(t)\rangle_{f(\Gamma,0)} = 0$, thus at sufficiently long times there is no dissipation. If we consider the ensemble average with respect to the distribution function at $t_c$, $f(\Gamma, t_c)$ and its resulting dissipation integral $\gamma\Delta g_{t_c}$, we see that $\gamma\langle \Delta g_{t_c}(\Gamma,t)\rangle_{f(\Gamma,t_c)} = 0$, and following the same arguments as above, the system can only be the dissipationless equilibrium state given by (12). This dissipationless distribution is unique for ergodic systems . We have therefore proven Boltzmann's postulate of equal *a priori* probabilities.

If the system is not ergodic the system relaxes to a number of possible quasi-equilibrium states. Such states break up phase space into ergodic subdomains within which the relative probabilities are Boltzmann distributed but between which, the relative weights are history dependent non-Boltzmann distributed [19]. The inter sub-domain weights at late times are history dependent and change for different initial distributions.